\documentclass[preprint,showpacs,showkeys,preprintnumbers,amsmath,amssymb]{revtex4}
\usepackage{graphicx}% Include figure files
\usepackage{dcolumn}% Align table columns on decimal point
\usepackage{bm}% bold math

\begin{document}
\def\be{\begin{equation}}
\def\ee{\end{equation}}
\def\bc{\begin{center}}
\def\ec{\end{center}}
\def\bea{\begin{eqnarray}}
\def\eea{\end{eqnarray}}
\def\veps{\varepsilon}
\def\vsg{\varsigma}
\def\eps{\epsilon}
\def\lmd{\lambda}
\def\ol{\overline}
\def\ul{\underline}
\def\wt{\widetilde}

\preprint{APS/123-QED}

\title{Level Curvature distribution \\ in a model of two uncoupled
chaotic subsystems}

\author{G\"{u}ler Erg\"{u}n}
\email{Guler.Ergun@brunel.ac.uk}
\author{Yan V. Fyodorov}%
 \email{Yan.Fyodorov@brunel.ac.uk}
\affiliation{Department of Mathematical Sciences,\\ Brunel
University, \\ Uxbridge, UB8 3PH, United Kingdom.}

\date{July 18 2003}% It is always \today, today,

\begin{abstract}
We study distributions of eigenvalue curvatures for a block
diagonal random matrix perturbed by a full random matrix. The most
natural physical realization of this model is a quantum chaotic
system with some inherent symmetry, such that its energy levels
form two independent subsequences, subject to a generic
perturbation which does not respect the symmetry. We describe
analytically a crossover in the form of a curvature distribution
with a tunable parameter namely the ratio of inter/intra subsystem
coupling strengths. We find that the peak value of the curvature
distribution is much more sensitive to the changes in this
parameter than the power law tail behaviour. This observation may
help to clarify some qualitative features of the curvature
distributions observed experimentally in acoustic resonances of
quartz blocks.

\end{abstract}


 \maketitle

%************************************************************
\section{Introduction}
\label{sec:intro} Studies of statistical properties of complex
quantum systems (chaotic or disordered) show that their eigenvalue
spectra  exhibit patterns of universal fluctuations, whose
structure mainly depends on the fundamental symmetries of the
Hamiltonian \cite{Haake,Guhr}. Such a universality opens an
attractive possibility of  modelling the fluctuations by comparing
them with those observed in long sequences of eigenvalues of
random matrices of appropriate symmetry \cite{dyson,mehta}.
Namely, systems with no time reversal invariance are known to be
adequately described by Gaussian Unitary Ensemble (GUE) of complex
Hermitian matrices, and systems with time reversal invariance are
described by Gaussian Orthogonal (GOE)  or Gaussian Symplectic
(GSE) ensembles of real symmetric or complex quaternion matrices,
depending on the existence of strong spin-orbit coupling.

More recently, the interest in studying spectral statistics of
quantum chaotic systems was much revitalized by the understanding
that their energy spectra display a universal change in their
characteristics as a response to external perturbations of various
kinds. The nature of the perturbation may vary considerably
depending on the physical system, and usually involves application
of external fields (magnetic or electric), change of boundary
conditions or the shape of the system, rearrangement of positions
of impurities in disordered medium, or variation of temperature,
pressure or any other tunable physical characteristics.

One of  the frequently used measures of parametric
sensitivities of complex quantum systems is the
distribution of level curvatures, which are defined as the second
order derivative of eigenvalues with respect to a perturbation
parameter.

In \cite{Gasp} Gaspard and co-authors developed expressions for
the probability densities of level curvatures and found that the
large curvatures must exhibit universal behaviour classified
according to the underlying gross symmetries. Indeed, the
curvatures become large in the vicinity of avoided crossings of
energy levels as functions of a parameter
 and their distribution can be simply related to that of small
eigenvalue spacings. On the basis of extensive numerical
investigations of both random matrices and several quantum chaotic
systems, Zakrzewski and Delande conjectured an analytical
expression for the full distribution of curvatures
\cite{Z-D-curv}. Later on Zakrzewski-Delande formulae were derived
analytically by von Oppen \cite{Oppen-curv} and by Fyodorov and
Sommers \cite{Fyod-curv} for the random matrix models of all three
universality classes.

However, in many relevant experimental circumstances, physical
systems have accidentally more underlying symmetries (frequently
called ``geometric") acting  in addition to the presence or the
absence of the time-reversal invariance. Such symmetries naturally
induce classification of energy levels according to irreducible
representations of the corresponding symmetry group, and the
energy levels corresponding to different representations form
statistically independent subsequences. Only these subsequences
may be then meaningfully compared with the universal random matrix
patterns. In fact, a generic situation may be even more
complicated, since geometric symmetries may not be exact, but
approximate. This will clearly lead to spectra being a mixture of
different subsequences with uncertain statistical consequences of
mutual interference.

Recent experimental studies on acoustic resonance spectra in
quartz blocks \cite{quartz,quartz-detail}  suggest that the system
may fall into the
latter category, and the deviations from
standard theoretical predictions of parametric correlations
may have their origin in remnant geometric symmetries. This
fact motivated several groups to investigate the effects of
partial symmetry breaking on the level curvatures
\cite{symm-break, symm-break2}. However, closed form
analytical expressions for curvature distributions
for the case of partly broken symmetries are not available, to the
best of our knowledge.

In the present paper we consider a simpler, but related model.
Rather than studying level curvatures in a system with partly
broken symmetry, we address the case of two non-interacting
subsystems subject to a perturbation which induces both coupling
between the subsystems and variation of parameters within each of
the two subsystems. This should be generically the case for a
perturbation which does not respect the underlying symmetry. Of
course, we do not claim that our simple model to be adequate for
describing experimental situation in quartz blocks, but we rather
hope that our results could help in indicating how various factors
may affect the shape of level curvature distributions.

Employing the random matrix calculations allows to derive exact
expressions for the level curvature distribution as a function of
relative weight of induced inter- and intra-sublattice osculating
variations. The curvature distribution naturally interpolate
between the simple Cauchy-Lorentz shape - characteristic for a
pure symmetry-breaking perturbations which couple two subsystems
without modifying them individually, - and the Zakrzewski-Delande
formulas typical for perturbations respecting the underlying
symmetry.

We first treat a simpler case of complex Hermitian matrices
(systems with  broken time reversal invariance) in detail, and
then extend the derivations to the case of real symmetric
matrices. Our analytical calculations are supported and
corroborated by accurate  numerical simulations of random matrix
ensembles.

%************************************************************
\section{General relations}
\label{sec:GUE-analytical} To study the parametric dependence of
energy levels of a system with some underlying symmetry, we
consider a random matrix model where the Hamiltonian ${\cal H}$ of
the system linearly depends on a perturbation parameter $\veps$:
\be\label{model}{\cal H}(\veps)=\hat{A}+\veps\hat{B}.\ee For the
unperturbed Hamiltonian $\hat{A}$ we choose a block-diagonal
matrix $\hat{A}=\left(\begin{array}{cc}\hat{H_1}&{\bf
\oslash}\\{\bf \oslash}&\hat{H_2}\end{array}\right)$, with
$\hat{H}_{1,2}$ being $N\times N$ random matrices (complex
Hermitian or real symmetric) taken either from GUE or,
respectively, from GOE. These can be thought of as representing
two non interacting chaotic subsystems. That means, we consider
the matrices $\hat{H}_p$ as random Gaussian, with entries
${H}_{i>j}$ being independent and identically distributed
variables with mean zero and variances
$<\ol{H}_{ij}H_{ij}>=2\sigma^2/N$ for the GUE case ($\beta=2$),
and $<{H}_{ij}^2>=\sigma^2/2N$ for GOE case ($\beta=1$). The joint
probability density for $\hat{H}_p$ is then written as
\be\label{probdh}{\mathcal{P}}(\hat{H_p})=C^{\beta}_N\,\exp\left\{-
\frac{N}{2\beta\sigma^2}Tr\hat{H_p}^2\right\},\ee where
$C^{\beta}_N$ is the appropriate normalization constant.

We introduce an interaction between blocks $\hat{H}_{1,2}$ by
considering a general coupling matrix $\hat{B}=
\left(\begin{array}{cc}\hat{J}_{1} & \hat{V} \\ \hat{V}^\dag &
\hat{J}_{2}\end{array}\right)$. We assume here that
$\hat{J}_{1,2}$ have the same symmetry properties as the diagonal
blocks $\hat{H}_{1,2}$ and their probability densities can be
written as \be\label{prob-jb} {\mathcal{P}}(\hat{J}_p)\propto
\exp\left\{-\frac{N}{2\beta\sigma^2_1}Tr\hat{J}_p^2\right\}.\ee As
for the off-diagonal blocks $\hat{V}$, they represent general
Gaussian random matrices (complex for $\beta=2$ or real for
$\beta=1$) with no further symmetry constraints are imposed. The
probability density for $\hat{V}$ is then chosen for both cases to
be \be\label{probdv}{\mathcal{P}}(\hat{V})\propto
\exp\left\{-\frac{N}{2\sigma_2^2}Tr\hat{V}^\dag
\hat{V}\right\}.\ee \noindent

Models of this kind were previously employed with satisfactory
results in the analysis of data relative to symmetry breaking in
nuclear physics \cite{nuc}.

Denote the eigenvalues and corresponding eigenvectors of
$\hat{H}_p$ by $[\lmd^{(p)}_i,{\bf v}^{(p)}_i]$ with $p=1,2.$ This
implies that $\hat{H}_p{\bf v}^{(p)}_i=\lmd^{(p)}_i{\bf
v}^{(p)}_i$ where $i=1,\ldots, N$ and ${\bf v}^{(p)\dag}_i{\bf
v}^{(p)}_i=1$. Our main goal is to find the distribution of level
curvatures, defined as the second order derivative of eigenvalues
of ${\cal H}$ with respect to the perturbation parameter $\veps$.
Employing in the usual way the second order perturbation theory we
can write the expression for the curvature corresponding, say, to
an eigenvalue that for $\veps=0$ coincides with the (unperturbed)
eigenvalue $\lambda_i^{(1)}$ as: \be\label{eq-l1i2}
{\mathcal{C}}_{i}=\sum_{k\neq i}^N
 \frac{({\bf v}^\dag_{k1}\hat{J}_1{\bf v}_{i1})
       ({\bf v}^\dag_{i1}\hat{J}^\dag_1{\bf v}_{k1})}
      {\lmd^{(1)}_i- \lmd^{(1)}_k}
+\sum_{k=1}^N\frac{({\bf v}^\dag_{i1}\hat{V}{\bf v}_{k2})
                   ({\bf v}^\dag_{k2}\hat{V}^\dag{\bf v}_{i1})}
      {\lmd^{(1)}_i- \lmd^{(2)}_k}.\ee
This expression shows that there are generically two contributions
to the level curvatures. The first sum is essentially the level
curvature induced by the {\it block-diagonal} part of the
perturbation which does not lead to any mixing between levels of
the two non-interacting subsystems. Taken alone, this term (which
we will denote here as ${\mathcal{C}}_{1i}$) must, therefore,
yield the Zakrzewski-Delande curvature distribution. In contrast,
the second sum that will be denoted as ${\mathcal{C}}_{2i}$
reflects the influence of the off-diagonal perturbation, which
mixes the levels of two subsystems.

Then the distribution of the total curvatures defined as
${\mathcal{P(C)}}=\left\langle
\delta\left({\mathcal{C}}-{\mathcal{C}}_{1i}- {\mathcal{C}}_{2i}
\right)\right\rangle_{{\cal H}}$ and it can be conveniently
written using the Fourier transform as
\be\label{curvdist2}{\mathcal{P(C)}}=
\frac{1}{2\pi}\int_{-\infty}^{\infty}dx
e^{ix{\mathcal{C}}}\left\langle \left\langle
e^{-ix{\mathcal{C}}_{1i}} \right\rangle_{\hat{J}_1}\left\langle
e^{-ix{\mathcal{C}}_{2i}}\right\rangle_{\hat{V}}
\right\rangle_{\hat{H}_1, \hat{H}_2}.\ee The first factor
$\left\langle \left\langle
e^{-ix{\mathcal{C}}_{1i}}\right\rangle_{\hat{J}_1}
\right\rangle_{\hat{H}_1}$ is just the Fourier transform of the
known Zakrzewski-Delande expression, hence the calculation of the
curvature distribution reduces to evaluating the remaining factor
$\left\langle\left\langle
e^{-ix{\mathcal{C}}_{2i}}\right\rangle_{\hat{V}}
\right\rangle_{\hat{H}_2}$. Our next goal is to derive the
corresponding expressions, first for $\beta=2$ and then for
$\beta=1$.
%In fact, we give an independent detailed derivation of Zakrzewski-Delande
%formulas for $\beta=1,2$ in the Appendix.

\section{\bf Complex Hermitian matrices: $\beta=2$}

We are going to evaluate the following ensemble average
(see Eq.(\ref{eq-l1i2}))
 \be \label{fcurvdist}\left\langle
e^{-ix\sum^N_{k=1}\frac{1}{\lmd^{(1)}_i-\lmd^{(2)}_k} ({\bf
v}^\dag_{i1}\hat{V}{\bf v}_{k2})({\bf v}^\dag_{k2}\hat{V}^\dag{\bf
v}_{i1})}\right\rangle_{\hat{V},\hat{H}_2}.\ee

First, we perform the average over $\hat{V}$. To simplify the
notation we denote $x/(\lmd^{(1)}_i-\lmd^{(2)}_k)\equiv b_k$,
${\bf v}^\dag_{i1}\hat{V}{\bf v}_{k2} \equiv w_k$  and use the
identity
\be\label{identity}e^{-ib_k\ol{w}_kw_k}=-\frac{i}{b_k}\int
\frac{d\ol{Z}_kdZ_k}{2\pi}e^{\frac{i}{b_k}\ol{Z}_kZ_k-i(Z_kw_k+\ol{Z}_k\ol{w}_k)},\ee
where the integration is taken over an  auxiliary complex variable
$Z_k$. This allows to rewrite Eq.(\ref{fcurvdist}) in the
following form:
\be\label{av-id}\left\langle\prod_{k=1}^N\left(-\frac{i}{
b_k}\right)\int
\frac{d\ol{Z}_kdZ_k}{2\pi}e^{\frac{i}{b_k}\ol{Z}_kZ_k-i(Z_kw_k+
\ol{Z}_k\ol{w}_k)}\right\rangle_{\hat{V}, \hat{H}_2}.\ee

Explicitly, employing the distribution function in
Eq.(\ref{probdv}), we need to calculate the integral
\be\label{avV} \int d\hat{V}d\hat{V}^\dag \exp
\left\{-\frac{N}{2\sigma_2^2}Tr(\hat{V}\hat{V}^\dag)- i{\bf
v}^\dag_{i1}\hat{V}\left(\sum_{k=1}^N Z_k{\bf v}_{k2} \right)
-i\left(\sum_{k=1}^N\ol{Z}_k{\bf
v}^\dag_{k2}\right)\hat{V}^\dag{\bf v}_{i1}\right\}.\ee

It is convenient to introduce an $N \times N$ matrix
$\hat{G}=\left(\sum_kZ_k{\bf v}_{k2}\right)\otimes{\bf
v}^\dag_{i1}$, so that, when used in the former identity relation
we get \be\label{gen-identity}\int d\hat{V}d\hat{V}^\dag
e^{-\frac{N}{2\sigma_2^2}Tr(\hat{V}\hat{V}^\dag)-iTr(\hat{V}
\hat{G}+\hat{G}^\dag \hat{V}^\dag)}\propto
e^{-\frac{2\sigma_2^2}{N}Tr(\hat{G}^\dag\hat{G})},\ee which is a
generalization of Eq.(\ref{identity}). We further use
\be\label{Tr}Tr(\hat{G}^\dag\hat{G})=({\bf v}^\dag_{i1}{\bf
v}_{i1})\sum_{k\varrho}\ol{Z}_k Z_\varrho({\bf v}^\dag_{k2}{\bf
v}_{\varrho2}),\ee and recall that ${\bf v}_{i1}$ are eigenvectors
of $\hat{H}_1$ and ${\bf v}_{k2}$ are those of $\hat{H}_2$. Using
the orthogonality of the eigenvectors: $$ {\bf v}^\dag_{k2}{\bf
v}_{\varrho2} = \left\{\begin{array}{l} 1, \quad k=\varrho \\0,
\quad k\neq\varrho\end{array}\right.$$ and  ${\bf v}^\dag_{i1}{\bf
v}_{i1}=1$, gives the result for the average in Eq.(\ref{av-id})
as \be \label{av-v} \left\langle\prod_{k=1}^N\int
\frac{d\ol{Z}_kdZ_k}{2\pi i
b_k}e^{-\frac{2\sigma_2^2}{N}\ol{Z}_kZ_k
+\frac{i}{b_k}\ol{Z}_kZ_k}\right\rangle_{\hat{H}_2}
\propto\left\langle\frac{\det\left(\lmd_i^{(1)}I_N-\hat{H}_2\right)}
{\det\left[\left(\lmd^{(1)}_i+\frac{i\wt{x}}{N}\right)I_N-\hat{H}_2\right]}
\right\rangle_{\hat{H}_2}.\ee For this we also performed the
Gaussian integrations over $Z_k$ explicitly, and denoted
$\wt{x}=2\sigma_2^2x$.

In what follows we denote $\lmd^{(1)}_i\equiv\lmd$ and
$\lmd^{(1)}_i+\frac{i\wt{x}}{N}\equiv\lmd_b$ and proceed with
calculations of the average
$\left\langle\cdots\right\rangle_{\hat{H}_2}$ in Eq.(\ref{av-v})
by employing a technique suggested in \cite{F-S}. In fact, for
$\beta=2$ the averages of the ratios of determinants are known in
full generality for any value of $N$ \cite{F-S, F-S1}.
Nevertheless, we outline the corresponding calculation in order to
introduce the method and the convenient notation which will be
used later on in this paper for the more complicated case
$\beta=1$.

Using the standard ``supersymmetrization" idea \cite{Efetov}, we
represent the denominator of the expression to be averaged as a
Gaussian integral (we assume here $x>0$ for definiteness)
\be\label{complex}{\det}^{-1}(\lmd_b
I_N-\hat{H}_2)=\frac{1}{i^N}\int d{\bf S}d{\bf S}^\dag
e^{\frac{i}{2}\lmd_b{\bf S}^\dag{\bf S}-\frac{i}{2}{\bf
S}^\dag\hat{H}_2{\bf S}},\ee  with a complex $N$ dimensional
vector ${\bf S}=(S_1,\ldots,S_N)^T$, where $T$ stands for the
vector transpose. For the determinant in the numerator, we use
Gaussian integrals over anticommuting (Grassmannian) $N$-
component vectors $\chi,\,\chi^\dag$, which gives
\be\label{grassman}\det(\lmd I_N-\hat{H}_2)=\frac{1}{i^N}\int
d\chi d\chi^\dag e^{\frac{i}{2}\lmd
\chi^\dag\chi-\frac{i}{2}\chi^\dag\hat{H}_2\chi}.\ee Substituting
the relations:
$\chi^\dag\hat{H}_2\chi=-Tr(\hat{H}_2\chi\otimes\chi^\dag)$ and
${\bf S}^\dag\hat{H}_2{\bf S}=Tr(\hat{H}_2{\bf S}\otimes{\bf
S}^\dag)$ in the integral, yields \be\label{av-new-form}
\left\langle\cdots\right\rangle_{\hat{H}_2}=\int d^2\chi\int
d^2{\bf S}e^{\frac{i}{2}[\lmd \chi^\dag\chi+\lmd_b{\bf S}^\dag{\bf
S}]}\left\langle e^{-\frac{i}{2}Tr\hat{H}_2[{\bf S}\otimes{\bf
S}^\dag-\chi\otimes\chi^\dag]}\right\rangle_{\hat{H}_2}.\ee The
ensemble average over GUE matrices $\hat{H}_2$ can be easily
performed by exploiting the identity \be\label{identitygue}
\left\langle
e^{-\frac{i}{2}Tr[\hat{H}\hat{A}]}\right\rangle_{GUE}\propto
e^{-\frac{\sigma^2}{4N}Tr[\hat{A}^2]}\ee and ``decoupling" the
quartic term in Grassmann variables with the help of simple
Hubbard-Stratonovich  transformation: \be\label{HS1}
e^{\frac{\sigma^2}{4N}(\chi^\dag\chi)^2}=\int_{-\infty}^{\infty}
\frac{dq}{\sqrt{2\pi}}e^{-\frac{q^2}{2}-\frac{q\sigma}{\sqrt{2N}}\chi^\dag\chi}.
\ee

After some straightforward manipulations we arrive at the
following integral representation for the required ensemble
average: \be\label{av-fin}
\left\langle\cdots\right\rangle_{\hat{H}_2}=\int d^2{\bf
S}e^{-\frac{\sigma^2}{4N}({\bf S}^\dag{\bf
S})^2+\frac{i}{2}\lmd_b{\bf S}^\dag{\bf S}}
\int^{\infty}_{-\infty}\frac{dq}{\sqrt{2\pi}}\,e^{-\frac{q^2}{2}}
\det\left[\frac{1}{2}(i\lmd-\frac{q\sigma}{N/2})\hat{I}_N-
\frac{{\bf S}\otimes{\bf S}^\dag}{2N/\sigma^2}\right].\ee

Introducing the variable $q_F=i\lmd-\frac{q\sigma}{\sqrt{N/2}} $
and shifting the contour of integration in such a way that, the
integral over $q_F$ goes along the real axis, we can rewrite the
above expression as \be\label{contour} \left\langle\cdots
\right\rangle_{\hat{H}_2}\propto
\int_{-\infty}^{\infty}\frac{dq_F}
{\sqrt{2\pi}}e^{-\frac{N}{4\sigma^2}(q_F-i\lmd)^2}\int d^2{\bf
S}e^{\frac{\sigma^2}{4N}({\bf S}^\dag{\bf S})^2
+\frac{i}{2}\lmd_b{\bf S}^\dag {\bf S}}
\det\left[q_F\hat{I}-\frac{\sigma^2}{N}{\bf S}\otimes{\bf
S}^\dag\right],\ee where we shifted the contour for $q_F
\in{(-\infty,\,\infty)}$ to be real. Further simplification can be
made by noticing that the $N\times N$ matrix ${\bf S}\otimes {\bf
S}^\dag$ is of rank unity, i.e. it has $(N-1)$ zero eigenvalues,
and only one nonzero eigenvalue equal to $({\bf S} {\bf S}^\dag)$.
Then the determinant in the previous expression is equal to
\be\label{det} \det\left[q_F\hat{I}-\frac{\sigma^2}{N}{\bf
S}\otimes{\bf S}^\dag\right]\equiv
q_F^{N-1}\left(q_F-\frac{\sigma^2}{N}{\bf S}{\bf
S}^\dag\right).\ee Finally, we introduce polar coordinates: ${\bf
S}=r{\bf n}$ with ${\bf n}^\dag{\bf n}=1$ and $\int d^2{\bf
S}=r^{2N-1}dr\, d{\bf n}$, where $\int d{\bf n}=\Omega_N$ produces
a constant factor, which corresponds to the area of a $2N$
dimensional unit sphere. Further introducing $p=r^2$ and changing
$p\to Np/\sigma^2$ and then following with the obvious
manipulations we get:
\bea\label{re-av}\left\langle\frac{\det(\lmd\hat{I}_N-\hat{H})}
{\det(\lmd_b\hat{I}_N-\hat{H})}\right\rangle_{\hat{H}_2} &=&
C_Ne^{\frac{N}
{4\sigma^2}\lmd^2}\int_{-\infty}^{\infty}\frac{dq_F}{\sqrt{2\pi}q_F}
e^{-\frac{N}{4\sigma^2}(q_F^2-2i\lmd q_F-4\sigma^2lnq_F)}\nonumber
\\&& \times \int_0^\infty
\frac{dp(q_F-p)}{\sqrt{2\pi}p}e^{-x\frac{\sigma_2^2}{\sigma^2}p}
e^{-\frac{N}{4\sigma^2}(p^2-2i\lmd p-4\sigma^2ln p)},\eea where we
reinstated $\lmd_b=\lmd+i{\wt x}/N$, ${\wt x}=2\sigma^2x$ and
$C_N$ stands for the accumulated constant factors. The latter can
always be restored by noticing that when $\lmd_b=\lambda$ the
right hand side must yield unity identically.

So far all the expressions were valid for finite-size matrices.
When $N \to \infty$ we expect the results, when appropriately
scaled, to be universal, i.e. broadly insensitive to the details
of the distribution of random matrices and applicable to quantum
chaotic systems. In such a limit the integrals in Eq.(\ref{re-av})
can be evaluated by the saddle point method. For
$\lmd\leq\sqrt{8\sigma^2}$ ( the so called \emph{bulk} of the
spectrum) the relevant saddle points are $p^{s.p}=
\frac{i\lmd+\sqrt{8\sigma^2-\lmd^2}}{2}$ and
$q_F^{s.p}=\frac{i\lmd\pm\sqrt{8\sigma^2-\lmd^2}}{2}$. It is easy
to see that only the choice $q_F^{s.p}=
\frac{i\lmd-\sqrt{8\sigma^2-\lmd^2}}{2}$ yields the leading-order
contribution, due to the presence of the factor $(q_F-p)$ in the
integrand. Substituting this choice into the integrand in
Eq.(\ref{re-av}) and evaluating the Gaussian fluctuations around
the saddle-point values, finally yields \be\label{av-finally}
\left.\left\langle\frac{\det(\lmd\hat{I}_N-\hat{H})}
{\det[(\lmd+i\frac{{\wt x}}{N})\hat{I}_N-\hat{H}]}
\right\rangle\right|_{^{x>0}_{N\to \infty}} =
\exp\left\{-\frac{x}{2}[i\lmd+\sqrt{8\sigma^2-\lmd^2}]
\left(\frac{\sigma_2}{\sigma}\right)^2\right\}. \ee It is easy to
repeat the calculation for $x<0$ and find that for any real value
of $x$ the result can be written as \be\label{x>0}
\left.\left\langle\frac{\det(\lmd\hat{I}_N-\hat{H})}
{\det[(\lmd+i\frac{{\wt x}}{N})\hat{I}_N-\hat{H}]}
\right\rangle\right|_{N\to\infty}=\exp \left\{-i\lmd
{x}\frac{\sigma_2^2}{2\sigma^2}-2|{x}|\pi
\rho(\lmd)\sigma_2^2\right\},\ee where
$\rho(\lmd)=\frac{1}{4\pi\sigma^2}\sqrt{8\sigma^2-\lmd^2}$ is the
mean eigenvalue density for GUE.

The Fourier-transform of the above expression with respect to $x$
immediately gives us the distribution
${\mathcal{P}}_{off}({\mathcal{C}})$ of level curvatures induced
by purely off-diagonal random coupling $\hat{V}$ between the two
subsystems. In the large-size limit ${N\to\infty}$ we therefore
have: \be\label{curv-dist1}
{\mathcal{P}}_{off}({\mathcal{C}})=\frac{1}{\pi}\frac{2\sigma_2^2\pi\rho(\lmd)}
{\left({\mathcal{C}}-\lmd\frac{\sigma_2^2}{2\sigma^2}\right)^2+
(2\pi\rho(\lmd)\sigma_2^2)^2},\ee which is nothing else but the
Cauchy-Lorentz distribution with the mean value $\langle
{\mathcal{C}}\rangle_{off}=\lmd\frac{\sigma_2^2}{2\sigma^2}$ and
characteristic widths $\Gamma_{off}=(2\pi\rho(\lmd)\sigma_2^2)$.

Turning our attention to the curvatures induced by the
block-diagonal contributions $\hat{J}_p$ (the term
${\mathcal{C}}_{1i}$ in Eq.(\ref{eq-l1i2}) we can  first perform
the ensemble average over the Gaussian distribution of
$\hat{J}_1$, Eq.(\ref{prob-jb}). Employing similar methods as
before, we easily find the result to be
 \be\label{av-J} \left\langle
e^{-ix{\mathcal{C}}_{1i}}\right\rangle_{\hat{J}_1}=\prod_{k\neq
i}\frac{(\lmd^{(1)}_i-\lmd^{(1)}_k)}{\left[\lmd^{(1)}_i-\frac{i
x_1} {N}- \lmd^{(1)}_k\right]},\ee where in this expression $
x_1=2\sigma_1^2 x$. This expression remains to be averaged over
the joint probability density of $N-1$ GUE eigenvalues
$\lambda^{(1)}_1,\ldots,\lambda^{(1)}_{i-1},
\lambda^{(1)}_{i+1},\ldots, \lambda^{(1)}_N$ which are different
from the chosen eigenvalue $\lambda^{(1)}_i$ whose curvature we
address. The consideration which is exposed
in\cite{Oppen-curv,Fyod-curv} shows that
\be\label{av1-wth}\left\langle
e^{-ix{\mathcal{C}}_{1i}}\right\rangle_{\hat{J}_1,\hat{H}_1}\propto
e^{-\frac{N}{4\sigma^2}\lmd^2}\left\langle\frac{\det^3(\lmd- H)}
{\det(\lmd+\frac{ix_1}{N}- H)}\right\rangle_{ N-1},\ee where $H$
is a $(N-1)\times(N-1)$ GUE matrix.

The averaging of the ratios of determinants in Eq.(\ref{av1-wth})
can be done, \emph{mutatis mutandis}, by the same
``supersymmetrization" procedure as above. The detailed exposition
of the corresponding calculation can be found in the paper by
Fyodorov and Strahov\cite{F-S}. Here we briefly sketch the main
steps.  After representing each of four determinants by the
Gaussian integrals (three over anticommuting and one over usual
complex variables) one can easily perform the GUE average by
exploiting the identity Eq.(\ref{identitygue}). Then the terms in
the exponent quartic with respect to anticommuting variables are
``decoupled" by introducing an auxiliary integration over $3\times
3$ Hermitian matrix $\hat{Q}$, the procedure being a
straightforward generalization of the Hubbard-Stratonovich
transformation (\ref{HS1}). All the subsequent manipulations are
quite analogous to those exposed above, and for our case, instead
of Eq.(\ref{re-av}) we arrive at its analogue pertinent: \bea
\label{av-st}&&\left\langle\frac{\det^3(\lmd-{\wt H})}
{\det(\lmd_b-{\wt H})}\right\rangle_{{\wt H}}\propto \int
dQ_F(\det Q_F)^{{\wt N}-1}\exp\left\{-\frac{{\wt
N}}{4\sigma^2}Tr(Q_F-i\lmd\hat{I})^2\right\}\nonumber \\ && \times
\int^\infty_0 dp\,p^{{\wt N}-1}e^{-x\frac{\sigma_2^2}{\sigma^2}p}
 \exp\left\{-\frac{{\wt
N}}{4\sigma^2}(p^2-2i\lmd\, p)\right\} \left(q^{(1)}_F-p\right)
\left(q^{(2)}_F-p\right) \left(q^{(3)}_F-p\right), \eea where
$q^{(1,2,3)}_F$ are real eigenvalues of the Hermitian $3\times 3$
matrix $Q_F$ and ${\wt N}$ stands for $N-1$.  In fact, since
$Tr(Q_F-i\lmd\hat{I})^2$ and $\det{Q_F}$ depend only on the
eigenvalues $q^{(1,2,3)}_F$, it is convenient to use these
eigenvalues and the corresponding eigenvectors as integration
variables. In these coordinates the integration measure is given
by \[ dQ_F\propto d\mu[U]dq^{(1)}_Fdq^{(2)}_Fdq^{(3)}_F\prod_{1\le
k_1< k_2\le 3}\left(q^{(k_1)}_F-q^{(k_2)}_F\right), \] where
$d\mu[U]$ is the invariant measure on the manifold of unitary
$3\times 3$ matrices, representing the eigenvectors of $Q_F$ and
the last factor is the Jacobian of the transformation, known as
the Vandermonde determinant.

Again, we are  interested in the limit $N\gg 1$, so we neglect the
difference between $N$ and $N-1$ and omit the tilde henceforth.
The set of the saddle points of the integrand with respect to each
of the variables $p>0$ and $q^{(1,2,3)}_F$ is:
$p^{s.p}=\frac{i\lmd+\sqrt{8\sigma^2-\lmd^2}}{2}$ and
$q_F^{s.p}=\frac{i\lmd\pm\sqrt{8\sigma^2-\lmd^2}}{2}$. These
saddle points are the same as what we found earlier. However, the
presence of both the Vandermonde factors and that of the product
$\prod_{k=1}^3\left(q^{(k)}_F-p\right)$ make us select the
following saddle points:
\[
q_F^{(1)}=\frac{i\lmd+\sqrt{8\sigma^2-\lmd^2}}{2}\quad,\quad
q_F^{(2)}=q_F^{(3)}=\frac{i\lmd-\sqrt{8\sigma^2-\lmd^2}}{2}
\]
(as well as its cyclic permutations) as these give the
leading-order contribution. In fact, the integrand vanishes at
these saddle-point values and care should be taken to expand the
integrand further when calculating the contribution from the
Gaussian fluctuations around the saddle-points (see \cite{F-S} for
a general procedure). The final result is given by
\be\label{avo-wth}\left\langle
e^{-ix{\mathcal{C}}_{1i}}\right\rangle_{\hat{J}_1,\hat{H}_1}=
\left[1+2\pi\rho(\lambda)\sigma_1^2|x|\right] \exp\left\{-i\lmd
{x}\frac{\sigma_1^2}{2\sigma^2}-2|{x}|\pi
\rho(\lmd)\sigma_1^2\right\}. \ee Taking the Fourier-transform,
we, as expected, arrive at the Zakrzewski-Delande formula for
$\beta=2$: \be\label{curv-dist2}
{\mathcal{P}}_{diag}({\mathcal{C}})=\frac{2}{\pi}\frac{\Gamma^3_d}
{\left[\left({\mathcal{C}}-\langle
{\mathcal{C}}\rangle_d\right)^2+\Gamma^2_d\right]^2},\ee where the
mean value $\langle
{\mathcal{C}}\rangle_{d}=\lmd\frac{\sigma_1^2}{2\sigma^2}$ and the
characteristic widths $\Gamma_{d}=(2\pi\rho(\lmd)\sigma_1^2)$.

Now we know all the factors in Eq.(\ref{curvdist2}) and can find
the curvature distribution accounting for both diagonal and
off-diagonal perturbations of the two decoupled subsystems:
\be\label{gen-case}
{\mathcal{P(C)}}=\int^\infty_{-\infty}\frac{dx}{2\pi}e^{ix{\mathcal{C}}}
\left[1+\Gamma_d|x|\right]\exp\left\{-ix
\frac{\lmd}{2\sigma^2}(\sigma_1^2+\sigma_2^2)-|x|
\left(\Gamma_{off}+\Gamma_d\right)\right\}.\ee Performing the
integration explicitly, we arrive at our final formula for complex
Hermitian case: \be\label{curfin}
{\mathcal{P(C)}}=\frac{1}{\pi}\left\{\frac{\Gamma_{off}}
{\left({\mathcal{C}}-\frac{1}{2}
(\langle{\mathcal{C}}\rangle_{off}+\langle
{\mathcal{C}}\rangle_d)\right)^2+
\left(\Gamma_{off}+\Gamma_d\right)^2}\right. \ee
\[
\left.+\frac{2\Gamma_d \left(\Gamma_{off}+\Gamma_d\right)^2}
{\left[\left({\mathcal{C}}-\frac{1}{2} (\langle
{\mathcal{C}}\rangle_{off}+\langle
{\mathcal{C}}\rangle_d)\right)^2+
\left(\Gamma_{off}+\Gamma_d\right)^2\right]^2} \right\}.\]

\section{\bf Real symmetric matrices: $\beta=1$}

We again need to evaluate the ensemble average as in
Eq.(\ref{fcurvdist}), but this time for the real-valued
perturbation $\hat{V}$ and real-valued eigenvectors ${\bf v}_i$,
so that the quantity ${\bf v}^T_{i1}\hat{V}{\bf v}_{k2} \equiv
w_k$ is a real variable. As before we denote
$x/(\lmd^{(1)}_i-\lmd^{(2)}_k)\equiv b_k$ and use the integration
over an auxiliary real variable $x_k$:
\be\label{identityo}e^{-ib_kw^2_k}=-\sqrt{\frac{i}{b_k}}\int
\frac{dx_k}{2\pi}e^{\frac{i}{4b_k} x^2_k-ix_k w_k},\ee combined
with the fact that (cf. Eqs. (\ref{gen-identity})) \be\label{avVo}
\int d\hat{V}\exp \left\{-\frac{N}{2\sigma_2^2}
Tr(\hat{V}\hat{V}^T)-iTr\hat{V}\sum_{k=1}^N x_k\left({\bf
v}_{k2}\otimes{\bf v}^T_{i1} \right)\right\}\propto\exp\left\{-
\frac{\sigma_2^2}{2N}\sum_{k=1}^Nx^2_k\right\} \ee because of the
orthogonality of eigenvectors. Consequently, we easily perform the
Gaussian integral over $x_k$ and obtain \be \label{av-vo}
\left\langle\left\langle
e^{-ix{\mathcal{C}}_{2i}}\right\rangle_V\right\rangle_{H_2}
\propto\left\langle\frac{\det^{1/2}\left(\lmd_i^{(1)}I_N-\hat{H}_2\right)}
{\det^{1/2}\left[\left(\lmd^{(1)}_i+\frac{i\wt{x}}{N}\right)
I_N-\hat{H}_2\right]} \right\rangle_{\hat{H}_2},\ee as before we
denoted $\wt{x}=2\sigma_2^2x$.

After denoting $\lmd^{(1)}_i\equiv\lmd$ and
 $\lmd^{(1)}_i+\frac{i\wt{x}}{N}\equiv\lmd_b$ for a less cumbersome
 expression, we then proceed with calculations of the average
$\left\langle\cdots\right\rangle_{\hat{H}_2}$ in Eq.(\ref{av-vo}).
To be able to employ the previous technique for $\beta=2$ case we
first rewrite: \be\label{ido} \frac{\det^{1/2}\left(\lmd
I_N-\hat{H}_2\right)} {\det^{1/2}\left(\lmd_b
I_N-\hat{H}_2\right)}\equiv \frac{\det\left(\lmd
I_N-\hat{H}_2\right)} {\det^{1/2}\left(\lmd_b I_N-\hat{H}_2\right)
\det^{1/2}\left(\lmd I_N-\hat{H}_2\right)}. \ee

Assuming, for definiteness, $\wt{x}<0$, and also assuming that
$\lambda$ has an infinitesimal negative imaginary part we can
represent the two factors in the denominator as Gaussian integrals
over real $N$- component vectors ${\bf x}_{1,2}$:
\be\label{real1}{\det}^{-1/2}(\lmd I_N-\hat{H}_2)\propto\int d{\bf
x}_1 e^{-\frac{i}{2}\lmd{\bf x}_1^T{\bf x}_1+\frac{i}{2}{\bf
x}_1^T\hat{H}_2{\bf x}_1}\ee and
\be\label{real2}{\det}^{-1/2}(\lmd_b I_N-\hat{H}_2)\propto\int
d{\bf x}_2 e^{-\frac{i}{2}\lmd_b{\bf x}_2^T{\bf
x}_2+\frac{i}{2}{\bf x}_2^T\hat{H}_2{\bf x}_2},\ee where $T$
stands for vector transpose. As for the determinant in the
numerator,  we can use the same Gaussian integral
Eq.(\ref{grassman}) over anticommuting (Grassmannian) $N$-
component vectors $\chi,\,\chi^\dag$. Substituting these integral
representations to Eq.(\ref{ido}) and performing the ensemble
averaging over GOE matrix $\hat{H}_2$, with the help of identity:
\be\label{identitygoe} \left\langle
e^{\pm\frac{i}{2}Tr[\hat{H}\hat{A}]}\right\rangle_{GOE}\propto
e^{-\frac{\sigma^2}{32 N} Tr\left(\hat{A}^T+\hat{A}\right)^2}, \ee
one can satisfy oneself that the resulting expression takes the
form: \be \int d{\bf x}_1d{\bf x}_2d\chi\, d\chi^\dag
e^{-\frac{i}{2}\left(\lmd{\bf x}_1^T{\bf x}_1+ \lmd_b{\bf
x}_2^T{\bf x}_2-\lambda\chi^\dag\chi\right)} \ee
\[
\times\exp\left\{-\frac{\sigma^2}{8N}Tr\left[\hat{Q}^2\right]
+\frac{\sigma^2}{16N}\left(\chi^\dag\chi\right)^2
+\frac{\sigma^2}{16N}\chi^\dag\left({\bf x}_1\otimes {\bf
x}_1^T+{\bf x}_2\otimes{\bf x}_2^T\right)\chi\right\}.
\]
In this expression we introduced a positive definite matrix
$\hat{Q}=\left(\begin{array}{cc}{\bf x}_1^T{\bf x}_1&{\bf
x}_1^T{\bf x}_2\\ {\bf x}_2^T{\bf x}_1 & {\bf x}_2^T{\bf x}_2
\end{array}\right)$.
Now we again use the  ``decoupling" of the quartic term in
Grassmann variables (the simple Hubbard-Stratonovich
transformation Eq.(\ref{HS1})) and then perform the Gaussian
integration over anticommuting variables explicitly. The latter
yields the determinant factor
\be\label{factor}\det\left[\left(i\lmd-\frac{q\sigma}{\sqrt{N}}\right)
\hat{I}_N+\frac{\sigma^2}{2N}\left({\bf x}_1\otimes {\bf
x}_1^T+{\bf x}_2\otimes{\bf x}_2^T\right) \right]. \ee This factor
can be brought to a simpler form \be\label{simp-fac}
\left(i\lmd-\frac{q\sigma}{\sqrt{N}}\right)^{N-2}
\det\left[\left(i\lmd-\frac{q\sigma}{\sqrt{N}}\right)\hat{I}_2+
\frac{\sigma^2}{2N}\left(\begin{array}{cc}{\bf x}_1^T{\bf
x}_1&{\bf x}_1^T{\bf x}_2\\ {\bf x}_2^T{\bf x}_1 & {\bf x}_2^T{\bf
x}_2 \end{array}\right) \right]\ee by noticing that $\left({\bf
x}_1\otimes {\bf x}_1^T+{\bf x}_2\otimes{\bf
x}_2^T\right)=\hat{X}\hat{X}^T$, where $\hat{X}=({\bf x}_1,{\bf
x}_2)$ is $N\times 2$ rectangular matrix, and using the identity:
$\det{(I_N-\hat{X}\hat{X}^T)}= \det{(I_2-\hat{X}^T\hat{X})}$ then
recognizing that $\hat{Q}$ introduced by us above is just $2\times
2$ matrix $\hat{X}^T\hat{X}$. We see that the resulting expression
depends on the vectors ${\bf x}_{1,2}$ only via the matrix
$\hat{Q}$. In recent papers \cite{F-S} it was shown that the
integration over ${\bf x}_{1,2}$ under these conditions can be
replaced by that over $\hat{Q}$, with an extra factor
$\det{Q}^{(N-3)/2}$ arising in the integration measure. After some
straightforward manipulations we arrive at the following integral
representation for the required ensemble average:
\bea\label{av-fino}
\left\langle\cdots\right\rangle_{\hat{H}_2}\propto
\int^{\infty}_{-\infty}\,dq\,e^{-q^2}
\left(i\lmd-\frac{q\sigma}{\sqrt{N}}\right)^{N-2}
\int_{Q>0}d\hat{Q}\det{Q}^{(N-3)/2}  \nonumber \\ \times
\exp\left\{-\frac{\sigma^2}{8N}Tr\left[\hat{Q}^2\right]-
\frac{i}{2}Tr\left[\hat{Q}\left(\begin{array}{cc}\lambda& 0 \\
0 & \lambda_b \end{array}\right) \right]\right\}
\det\left[\left(i\lmd-\frac{q\sigma}{\sqrt{N}}\right)\hat{I}_2+
\frac{\sigma^2}{2N}\hat{Q}\right]. \eea

Introducing the variables $q_F=-i\lmd+\frac{q\sigma}{\sqrt{N}}$
and $\hat{Q}_b=\frac{\sigma^2}{2N}\hat{Q}$ and shifting the
contour of integration in such a way that the integral over $q_F$
goes along the real axis, we can rewrite the above expression as
\bea\label{av-fino1}
\left\langle\cdots\right\rangle_{\hat{H}_2}\propto
\int^{\infty}_{-\infty}\,dq_F\,q_F^{N-2}
e^{-\frac{N}{\sigma^2}(q_F+i\lmd)^2}
\int_{Q_b>0}d\hat{Q}_b\det{Q_b}^{(N-3)/2}
\det\left[-q_F\hat{I}_2+\hat{Q}_b\right] \nonumber \\\times
\exp\left\{-\frac{N}{2\sigma^2}Tr\left[\hat{Q}_b^2\right]-
\frac{iN}{\sigma^2}Tr\left[\hat{Q}_b\left(\begin{array}{cc}\lambda& 0 \\
0 & \lambda_b \end{array}\right) \right]\right\}.\eea At the next
step we introduce appropriate polar coordinates in the space of
matrices $Q_b>0$: \be\label{polar}
\hat{Q}_b={\cal O}^T\left(\begin{array}{cc}p_1& 0 \\
0 & p_2 \end{array}\right) {\cal O}\quad,\quad d\hat{Q}_b\propto
|p_1-p_2|dp_1dp_2d{\cal O} \ee where $p_{1,2}>0$ and ${\cal O}$
are $2\times 2$ real orthogonal matrices: ${\cal O}^T{\cal
O}=I_2$, with $d{\cal O}$ being the corresponding Haar's measure.
Explicitly, we can parameterize
${\cal O}=\left(\begin{array}{cc}\cos{\phi}& \sin{\phi} \\
-\sin{\phi} & \cos{\phi} \end{array}\right)$ and $d{\cal
O}=d\phi/(2\pi)$. Substituting these expressions into
Eq.(\ref{av-fino1}) and after obvious manipulations we get \bea
\label{re-avo}
&&\left\langle\frac{\det^{1/2}(\lmd\hat{I}_N-\hat{H})}
{\det^{1/2}(\lmd_b\hat{I}_N-\hat{H})}\right\rangle_{\hat{H}_2} =
C_N e^{\frac{N}{\sigma^2}\lmd^2}
\int_{-\infty}^{\infty}\frac{dq_F}{q^2_F}
e^{-\frac{N}{\sigma^2}(q_F^2+2i\lmd
q_F-\sigma^2\ln{q_F})}\nonumber
\\&& \times \int_0^\infty\,dp_1\int_0^\infty\,dp_2
\frac{|p_1-p_2|}{(p_1p_2)^{3/2}}(q_F-p_1)(q_F-p_2) {\cal
I}_x(p_1,p_2) e^{-\frac{N}{2\sigma^2}\left({\cal L}(p_1)+{\cal
L}(p_2)\right)}, \eea where we reinstated $\lmd_b=\lmd+i{\wt
x}/N$, ${\wt x}=2\sigma_2^2 x$ and
\[
{\cal L}(p)=p^2+2i\lmd p-\sigma^2\ln{p}\quad,\quad
{\cal I}_x(p_1,p_2)=\int d\phi\,
e^{x\frac{\sigma_2^2}{\sigma^2}[(p_1+p_2)-(p_1-p_2)\cos{2\phi}]}
\]
and, as before, $C_N$ stands for the accumulated constant factors.
So far all expressions were valid for finite-size matrices. When
$N \to \infty$  the integrals in Eq.(\ref{re-avo}) are evaluated
by the saddle point method. For the bulk of the GOE spectrum
$\lmd\leq\sqrt{2\sigma^2}$, the relevant saddle points are
$p_{1,2}^{s.p}= \frac{-i\lmd+\sqrt{2\sigma^2-\lmd^2}}{2}$ and
$q_F^{s.p}=\frac{-i\lmd\pm\sqrt{2\sigma^2-\lmd^2}}{2}$. Again only
the choice $q_F^{s.p}= \frac{i\lmd+\sqrt{2\sigma^2-\lmd^2}}{2}$
yields the leading-order contribution, due to presence of the
factors $(q_F-p_{1,2})$ in the integrand. Substituting this choice
into the integrand in Eq.(\ref{re-avo}) and evaluating the
Gaussian fluctuations around the saddle-point values finally
yields \be\label{avo-finally}
\left.\left\langle\frac{\det^{1/2}(\lmd\hat{I}_N-\hat{H})}
{\det^{1/2}[(\lmd+i\frac{{\wt x}}{N})\hat{I}_N-\hat{H}]}
\right\rangle\right|_{^{x<0}_{N\to \infty}} =
\exp\left\{x[-i\lmd+\sqrt{2\sigma^2-\lmd^2}]
\left(\frac{\sigma_2}{\sigma}\right)^2\right\}. \ee It is easy to
repeat the calculation for $x>0$ and find that for any real value
of $x$ the result can be written as \be\label{1x>0}
\left.\left\langle\frac{\det^{1/2}(\lmd\hat{I}_N-\hat{H})}
{\det^{1/2}[(\lmd+i\frac{{\wt x}}{N})\hat{I}_N-\hat{H}]}
\right\rangle\right|_{N\to\infty}=\exp \left\{-i\lmd
{x}\frac{\sigma_2^2}{\sigma^2}-|{x}|\pi
\rho(\lmd)\sigma_2^2\right\},\ee where
$\rho(\lmd)=\frac{1}{\pi\sigma^2}\sqrt{2\sigma^2-\lmd^2}$ is the
mean eigenvalue density for GOE.

We see that in the large-size limit ${N\to\infty}$ the expression
for the level curvature distribution induced by purely
off-diagonal random coupling $\hat{V}$ between the two subsystems
is essentially the same Cauchy-Lorentz distribution for both
$\beta=2$ and $\beta=1$ cases, up to re-scaling of the widths and
the mean value with a simple factor 2: \be\label{curv-dist1o}
{\mathcal{P}}_{off}({\mathcal{C}})=\frac{1}{\pi}\frac{\sigma_2^2\pi\rho(\lmd)}
{\left({\mathcal{C}}-\lmd\frac{\sigma_2^2}{\sigma^2}\right)^2+
(\pi\rho(\lmd)\sigma_2^2)^2}.\ee We give a more detailed
discussion of this issue in the next section.

The distribution of curvatures induced by the block-diagonal
contributions $\hat{J}_p$ (the term ${\mathcal{C}}_{1i}$ in
\ref{eq-l1i2}) for real symmetric  matrices is quite different
from that of complex Hermitian ones. Performing the ensemble
average over the Gaussian distribution of
 $\hat{J}_1$, Eq.(\ref{probdv}) and employing the same methods
 we find the result to be
\ \be\label{avo-J} \left\langle
e^{-ix{\mathcal{C}}_{1i}}\right\rangle_{\hat{J}_1}=\prod_{k\neq
i}\frac{(\lmd^{(1)}_i-\lmd^{(1)}_k)^{1/2}}{\left[\lmd^{(1)}_i+\frac{i
x_1}{N}- \lmd^{(1)}_k\right]^{1/2}}\ee with $ x_1=2\sigma_1^2 x$.
The averaging over the joint probability density of $(N-1)$ GOE
eigenvalues $\lambda^{(1)}_1,\ldots,\lambda^{(1)}_{i-1},
\lambda^{(1)}_{i+1},\ldots, \lambda^{(1)}_N$ which are different
from the eigenvalue $\lambda^{(1)}_i$ whose curvature we address,
shows that \cite{Oppen-curv,Fyod-curv}
\be\label{avo-wth1}\left\langle
e^{-ix{\mathcal{C}}_{1i}}\right\rangle_{\hat{J}_1,\hat{H}_1}\propto
e^{-\frac{N}{2\sigma^2}\lmd^2}\left\langle\frac{|\det(\lmd-
H)|\,\det^{1/2}(\lmd-H)} {\det^{1/2}(\lmd+\frac{ix_1}{N}-
H)}\right\rangle_{ N-1},\ee where $H$ is a $(N-1)\times(N-1)$ GOE
matrix.

The averaging of the ratios of determinants in Eq.(\ref{avo-wth1})
can be done, \emph{mutatis mutandis}, by the same technique as
above. However, the presence of the absolute value of the
determinant makes accurate calculation to be quite lengthy, and it
will be presented elsewhere, but the result is compact and it is
given by \cite{Oppen-curv,Fyod-curv} \be\label{av-wth}\left\langle
e^{-ix{\mathcal{C}}_{1i}}\right\rangle_{\hat{J}_1,\hat{H}_1}=\pi\rho(\lmd)
\sigma_1^2|x|e^{-i\lmd {x}\frac{\sigma_1^2}{\sigma^2}}
K_1\left(|{x}|\pi\rho(\lmd)\sigma_1^2\right), \ee with $K_1(z)$
being the MacDonald function of the order one. Such an expression
yields, after the Fourier-transform, the Zakrzewski-Delande
formula for $\beta=1$: \be\label{curvo-dist2}
{\mathcal{P}}_{diag}({\mathcal{C}})=\frac{1}{2}\frac{\Gamma^2_d}
{\left[\left({\mathcal{C}}-\langle
{\mathcal{C}}\rangle_d\right)^2+\Gamma^2_d\right]^{3/2}}, \ee
where the mean value $\langle
{\mathcal{C}}\rangle_{d}=-\lmd\frac{\sigma_1^2}{\sigma^2}$ and the
characteristic widths $\Gamma_{d}=\pi\rho(\lmd)\sigma_1^2$.

The curvature distribution, accounting for both the diagonal and
the off-diagonal perturbations of two decoupled subsystems, can be
found as the convolution of two distributions:
\be\label{gen-caseo}
{\mathcal{P(C)}}=\int^\infty_{-\infty}d{\mathcal{C}}_1{\mathcal{P}}_{diag}({\mathcal{C}}_1)
{\mathcal{P}}_{off}({\mathcal{C}}-{\mathcal{C}}_1) \ee and in this
way we arrive at the final formula for the real symmetric case. We
present it below for the central point of the spectrum
$\lambda=0$: \be\label{curfino}
{\mathcal{P(C)}}=\frac{1}{2\pi}\int^\infty_{-\infty}d{\mathcal{C}}_1
\frac{\Gamma^2_{d}}{\left[{\mathcal{C}}_1^2+\Gamma_{d}^2\right]^{3/2}}
\frac{\Gamma_{off}}{\left({\mathcal{C}}-{\mathcal{C}}_1\right)^2+\Gamma_{off}^2}.
\ee

%**********************************************************

\section{Numerical Results and Discussions}
\label{sec:numerics}

In the present section we compare the derived analytical form of
the curvature distribution with the results of direct numerical
simulations of the ensemble. For our numerical investigations we
used a normal random distribution that was adopted from FORTRAN
Numerical Recipes \cite{N-recipes} and to find the eigenvalues we
superseded some subroutines from LAPACK \cite{lapack}.  To avoid
the necessity of unfolding the spectra we took into account only
levels around the central part of the spectrum. Namely, for a
$100\times100$ matrix, ten middle eigenvalues (20 or more for
larger matrices) were considered at each time step, and a
curvature value for each eigenvalue was calculated by a second
difference equation \be\label{sec-difference}
\lmd^{''}_i(\veps)\left|_{\veps=0}\right.=\frac{\lmd_i(-2\veps)+
16\lmd_i(-\veps)-30\lmd_i(0)+16\lmd_i(\veps)-\lmd_i(2\veps)}
{12\veps^2}.\ee The choice of five points instead of the usual
three \cite{Gasp} was made  to ensure the stability of the
results, especially for the GOE case of our system. The empirical
choice of $\veps=0.001$ was an outcome of a number of trials; the
values it takes may be system specific. We finally remark that
using larger matrices e.g. $400\times400$ did not improve the
quality of plots considerably.

\begin{figure}
%\centering \centerline{\epsfig{file=gue_lev.eps,width=10cm}}
\begin{center}
\includegraphics[scale=0.80,angle=0]{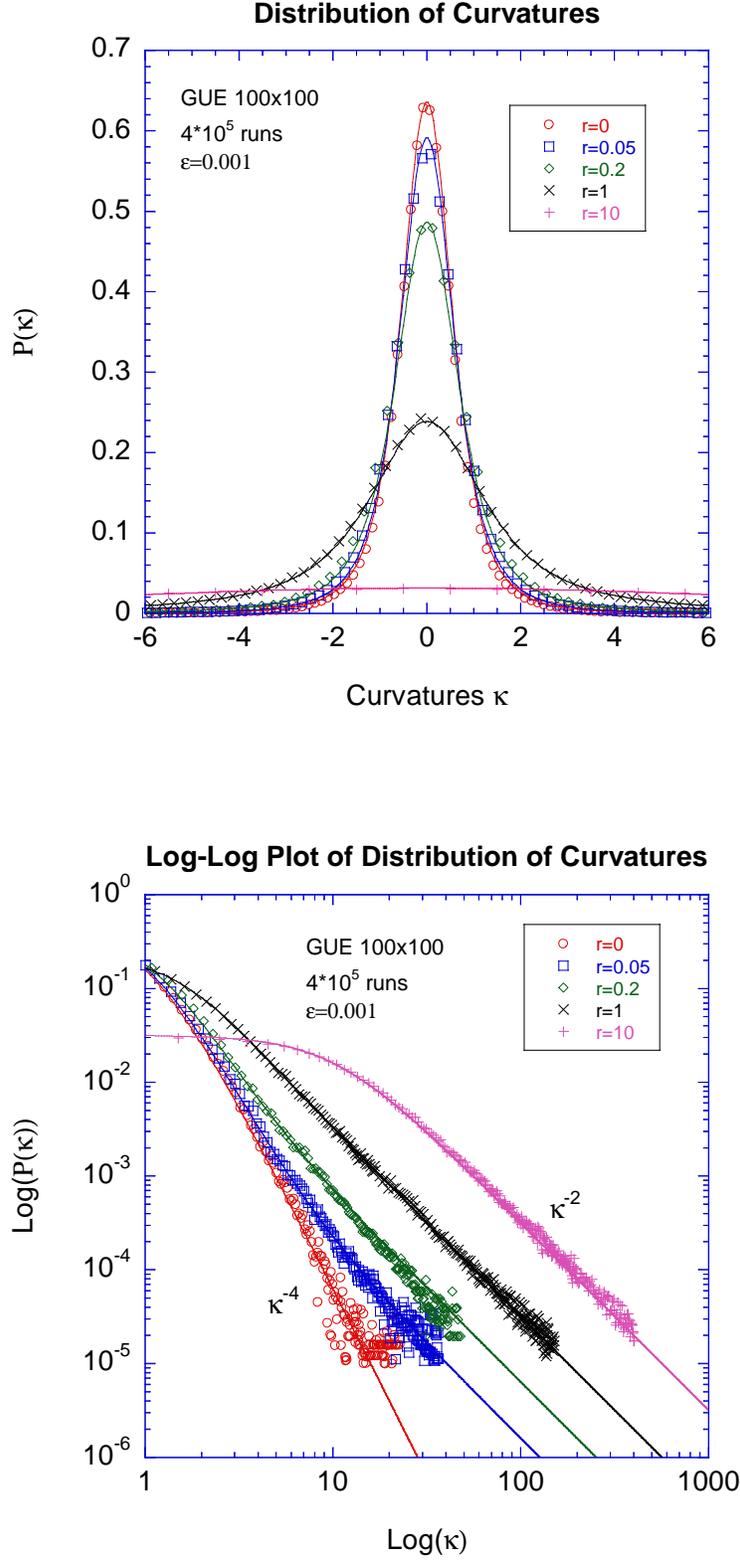}
\caption{\label{gue-inter} \noindent {Normalized GUE curvature
distributions for a few selected values of
$r=\sigma_2^2/\sigma_1^2$ (inter-coupling to intra-coupling
strengths);
 bottom plot (Log-Log scale) shows the tail behaviour of
 the same distribution.}}
\end{center}
\end{figure}

\begin{figure}
\begin{center}
%\centering \centerline{\epsfig{file=goe_lev.eps,width=10cm}}
\includegraphics[scale=0.80,angle=0]{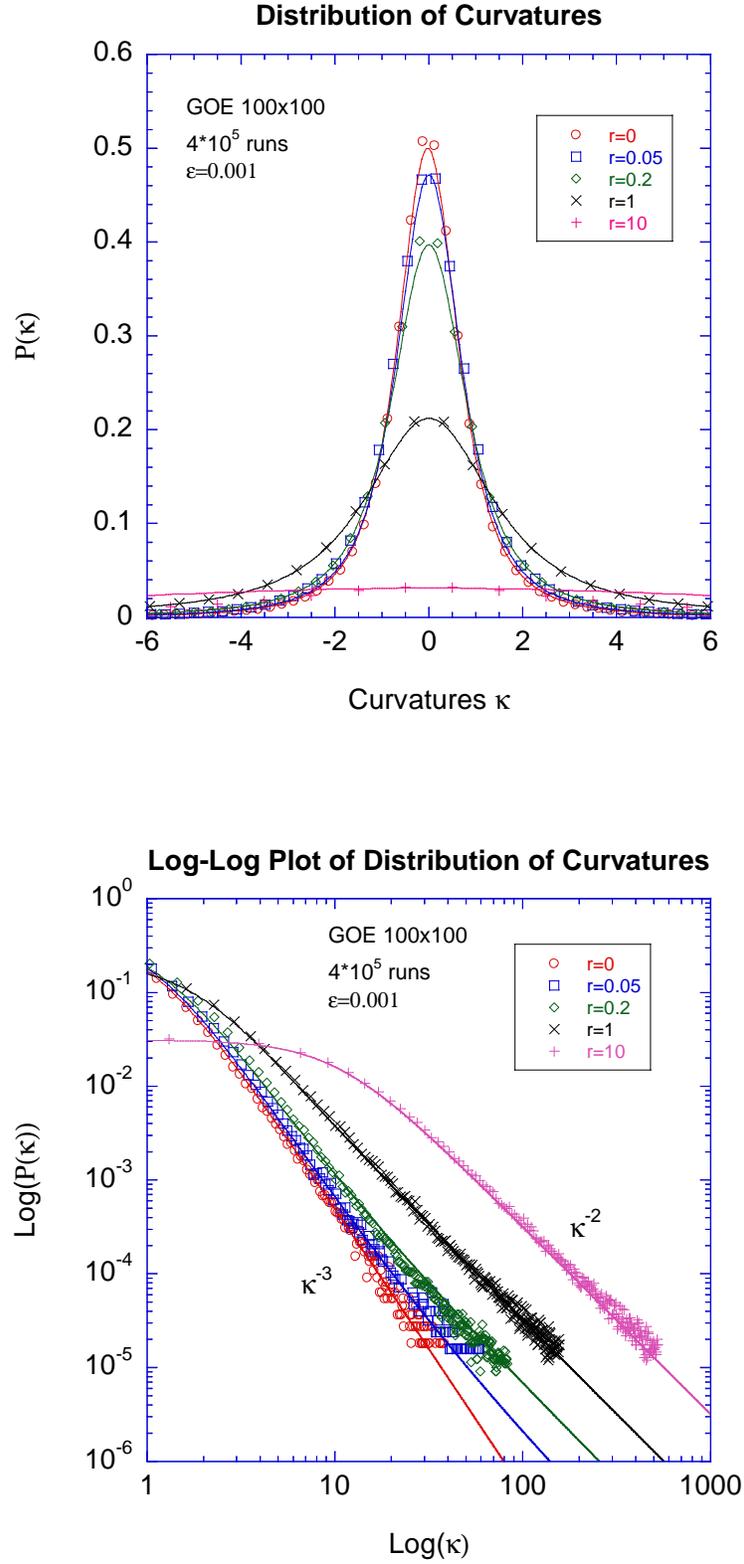}
\caption{\label{goe-inter} \noindent{Normalized GOE curvature
distributions for a few selected values of
$r=\sigma_2^2/\sigma_1^2$. The solid lines are analytical
predictions superimposed over numerical data plots.}}
\end{center}
\end{figure}

The normalized results of the simulations are presented in
Figs.(\ref{gue-inter},\ref{goe-inter}). To compare them with the
analytical predictions, for the GUE-like case $\beta=2$ we
consider Eq.(\ref{curfin}) at $\lmd=0$ and $\sigma=1$. It is also
convenient to use the dimensionless curvatures $\kappa$ obtained
from ${\cal C}$ by rescaling the latter with the variance of
``level velocity" as (cf. \cite{Oppen-curv}): \be \kappa={\cal
C}\frac{\Delta}{\pi\left\langle
\left(\frac{d\lambda^{(p)}_i}{d\veps} \right)^2\right\rangle} \ee
where $\Delta=\sigma\pi\sqrt{2}/N$ is the mean level spacing of a
single subsystem at $\lmd=0$. After some simple calculation, the
first order perturbation theory gives  $\left\langle
\left(\frac{d\lambda^{(p)}_i}{d\veps}\right)^2\right\rangle=
2\sigma^2\sigma_1^2/N$ and is independent of inter-subsystem
coupling strength $\sigma_2$. As a result of the rescaling, the
curvature distribution acquires the form \be\label{curfin-red}
{\mathcal{P(\kappa)}}=\frac{2}{\pi}
\left\{\frac{r/2}{\left[{\mathcal{\kappa}}^2+(1+r)^2 \right]}+
\frac{(1+r)^2}{\left[{\mathcal{\kappa}}^2+(1+r)^2\right]^2}\right\},\ee
controlled by the only parameter $r=\sigma_2^2/\sigma_1^2$ i.e the
ratio of the inter-subsystem to the intra-subsystem coupling
strengths. Superimposing the plots of this expression over the
appropriately normalized numerical data shows good agreement for
all corresponding values of the parameter $r$, despite the noise
in the large curvature tails. For curvatures exceeding the typical
value ${\mathcal{\kappa}}\gg r+1$ the distribution shows a power
law tail, the GUE-like behaviour ${\mathcal{\kappa}}^{-4}$ being
replaced by the Cauchy-Lorentz one ${\mathcal{\kappa}}^{-2}$ with
the relative growth of the ratio $r$. For any $r>0$ the most
distant tail is always of the Cauchy-Lorentz type, but
intermediate GUE-like behaviour is clearly seen for $r\ll 1$ when
intra-subsystem coupling appreciably exceeds the inter-subsystem
one. The crossover curvature value between the two regimes of
decay is approximately described by the expression \be\label{cc}
{\mathcal{\kappa}}_{cr}\simeq
\sqrt{2}\,\left(\frac{1}{\sqrt{r}}+\sqrt{r}\right), \ee which can
be obtained by equating the large-curvature tails originating from
the two competing terms in the expression Eq.(\ref{curfin-red}).
For curvatures in the interval $1\ll {\mathcal{\kappa}} \ll
{\mathcal{\kappa}}_{cr} \simeq \sqrt{2/r}$, the behaviour is
GUE-like, changing to a slower Cauchy-Lorentz decay at
${\mathcal{\kappa}}\gg {\mathcal{\kappa}}_{cr} \simeq \sqrt{2/r}$.
It is also easy to verify that the maximal value of the
distribution $P_{max}={\mathcal{P}}\left(0\right)$ always
decreases with the increasing ratio \be \label{peak}
{\mathcal{P}}\left(0\right)=\frac{1}{\pi}
\left[\frac{1}{r+1}+\frac{1}{(r+1)^2}\right] \ee

For the case of GOE, the similar rescaling of curvatures
 at $\lmd=0$ and $\sigma=\sqrt{2}$ leads from Eq.(\ref{curfino})
to the expression:
\be
{\cal P}(\kappa)=\frac{1}{2\pi}\int_{-\infty}^{\infty}
dx \frac{1}{(x^2+1)^{3/2}}\,\frac{r}{r^2+(x-\kappa)^2}
\ee

We plot this distribution superimposed over the numerical data for
various values of $r$. Again, they agree rather well with the
numerics and a crossover behaviour from GOE-like tail to
Cauchy-Lorentz one can be seen clearly for small $r$, e.g.
$r=0.05$.

In fact, one may notice from our plots that decrease in maximal
value of the distribution starts to be noticeable at much smaller
values of $r$ than the modification of tail behaviour at not very
large values of $\kappa$. This fact qualitatively corroborates
with the experimental observations in quartz blocks \cite{quartz},
where noticeable deviations were detected in the center of the
distribution, whereas the tails agreed well. Although, our
oversimplified model clearly can not be considered as adequate for
describing the actual experimental situation, we nevertheless
mention that the choice of $r \simeq 0.2$ allows matching the drop
of the peak value with the experimentally observed deviation and
produces an overall good agreement with the experimental curve.

As we already noted in the text of the paper, and as clearly seen
from the numerical Log-Log plots, the limiting large $-N$
curvature distribution due to purely off-diagonal
(inter-subsystem) perturbations turned out to have the same
Cauchy-Lorentz form irrespective of the underlying symmetry, for
both $\beta=2$ and $\beta=1$. Below we give an alternative,
heuristic derivation of this fact, which sheds some light on the
origin of such a behaviour. The starting point for our analysis
 is  expression Eq.(\ref{av-v}) for $\beta=2$ or
Eq.(\ref{av-vo}) for $\beta=1$. Denoting $\lmd_i^{(1)}=\lambda$,
as in the text above, we rewrite those formulas as: \bea
\label{av-v1} \left\langle\prod_{k=1}^N
\left[\frac{\left(\lmd-\lambda_{2,k}\right)}
{\left(\lmd+\frac{i\wt{x}}{N}\right)-\lambda_{2,k}}\right]^{\frac{\beta}{2}}
\right\rangle_{\hat{H}_2}= \left\langle\prod_{k=1}^N\left[1-
\frac{i\wt{x}}{N}\frac{1}
{\left(\lmd+\frac{i\wt{x}}{N}\right)-\lambda_{2,k}}\right]^{\frac{\beta}{2}}
\right\rangle_{\hat{H}_2} \nonumber \\
=\left\langle\exp\left\{{\frac{\beta}{2}}\sum_{k=1}^N
\log{\left[1- \frac{i\wt{x}}{N}\frac{1}
{\left(\lmd+\frac{i\wt{x}}{N}\right)-\lambda_{2,k}}\right]}\right\}
\right\rangle_{\hat{H}_2}\approx
\left\langle\exp{-{\frac{\beta}{2}}\sum_{k=1}^N
\frac{i\wt{x}}{N}\frac{1}
{\left(\lmd+\frac{i\wt{x}}{N}\right)-\lambda_{2,k}}}
\right\rangle_{\hat{H}_2}. \eea At the last step we made a
plausible assumption that the expression above in the limit of
large $N$ can be approximated by expanding the logarithms in the
exponential to the first non-vanishing term. Now we introduce an
{\it exact} eigenvalue density for $\hat{H}_2$ as: \be \label{dos}
\rho(\mu)=\frac{1}{N}\sum_{k=1}^N\delta(\mu-\lambda_{2,k}), \ee
where $\delta(x)$ stands for the Dirac delta function. In terms of
this density the ensemble average (\ref{av-v1}) can be rewritten
as: \be\label{av}
\left\langle\exp\left\{-i{\frac{\beta}{2}}\wt{x}\int
d\mu\rho(\mu)\frac{1}
{\left(\lmd+\frac{i\wt{x}}{N}\right)-\mu}\right\}
\right\rangle_{\hat{H}_2}. \ee Now, we use the well-known fact
that the exact eigenvalue density for random matrices is {\it
self-averaging}, which means in the limit $N\to \infty$ converges
to a {\it non-random} smooth function -the mean eigenvalue
density.
 The latter function is just given by the Wigner semicircular law
$\rho_{sc}(\mu)=\frac{2}{\pi\mu_{sc}^2}\sqrt{\mu_{sc}^2-\mu^2}$ for
$|\mu|<\mu_{sc}$, where $\mu_{sc}=\sqrt{8\sigma^2}$ for GUE and
$\mu_{sc}=\sqrt{2\sigma^2}$ for GOE.
 All these facts suggest that in the limit of large $N$ the
ensemble average in (\ref{av}) can be suppressed in favour of
replacing the exact density with its semicircular form. Moreover,
since $\rho_{sc}(\mu)$ is a smooth function, in the limit of $N\to
\infty$ we can use the Sohotsky formula: \be\label{sohotsky}
\lim_{N\to \infty}\int_{-\mu_{sc}}^{\mu_{sc}} d\mu\rho_{sc}(\mu)
\frac{1}{\left(\lmd+\frac{i\wt{x}}{N}\right)-\mu}= {\cal
P}\int_{-\mu_{sc}}^{\mu_{sc}} d\mu\rho_{sc}(\mu)\frac{1}
{(\lmd-\mu)}-isgn[\wt{x}]\pi\rho_{sc}(\lmd), \ee where the first
term is understood as a principal value integral, and $sgn$ stands
for the sign function of the argument. In fact, with some effort
the integral can be evaluated explicitly: \be\label{exp-int}
\frac{2}{\pi\mu_{sc}^2}{\cal P}\int_{-\mu_{sc}}^{\mu_{sc}}
 d\mu \frac{\sqrt{\mu_{sc}^2-\mu^2}}
{(\lmd-\mu)}=2\frac{\lambda}{\mu_{sc}^2}.\ee Collecting all terms
we see that: \be\label{scs}
\left.\left\langle\left[\frac{\det(\lmd\hat{I}_N-\hat{H})}
{\det[(\lmd+i\frac{{\wt
x}}{N})\hat{I}_N-\hat{H}]}\right]^{\beta/2}
\right\rangle\right|_{N\to\infty}=\exp
\left\{-\frac{\beta}{2}\left[i\wt{x}\frac{2\lmd}{\mu_{sc}^2}+|\wt{x}|\pi
\rho_{sc}(\lmd)\right]\right\}, \ee which coincides with the
earlier derived expressions in Eq.(\ref{x>0}) and in
Eq.(\ref{1x>0}).

It is natural to expect that such a derivation can be made
mathematically rigorous. However, despite its simplicity and
conceptual clarity, such a method can not be straightforwardly
applied to evaluation of the more complicated averages such as
those in Eqs.(\ref{av1-wth}) and (\ref{avo-wth1}). Indeed, the
application of the outlined procedure to Eq.(\ref{av1-wth})
amounts to approximating the extra determinant factor in the
large-N limit as: \be\label{large-lim}
{\det}^2{\left(\lambda\hat{I}_N-H_2\right)}\approx
\exp\left\{2N\frac{2}{\pi\mu_{sc}^2}{\cal
P}\int_{-\mu_{sc}}^{\mu_{sc}} d\mu
\sqrt{\mu_{sc}^2-\mu^2}\log{(\lambda-\mu)}\right\}=
e^{\frac{2N\lambda^2}{\mu_{sc}^2}},\ee which is indeed a correct
expression up to  the {\it leading} order in $N$ in the
exponential. It serves to cancel the extra factor
$e^{-\frac{N\lambda^2}{4\sigma^2}}$ in front of the ensemble
average in Eq.(\ref{av1-wth}). However, it is easy to understand
that to arrive to the correct expression Eq.(\ref{avo-wth}) one
needs to take into account {\it subleading} terms -those of the
order of unity in the exponential. This goal goes beyond the
simple use of self-averaging, and requires a much more detailed
treatment. It is not clear at the moment how to implement such a
treatment in the present heuristic scheme. That is why the
``supersymmetrization" method, which yields fully controllable
results in all cases should be in general preferred.

In conclusion, we have derived exact expressions for the
distribution of level curvatures in a model describing a mixing of
two independent spectra by a generic perturbation. Although the
model is too simple to describe actual experimental situation in
systems with partially broken symmetries, some features of the
behaviour of our curvature distribution may play a role of useful
analogy helping to understand the deviations in experimentally
measured level curvature distribution of the acoustic resonances
of quartz blocks \cite{quartz}. Indeed, the maximum value of the
latter distribution was found to be considerably lower than
predicted for pure GOE case, whereas the tail shows good
${\mathcal{C}}^{-3}$ decay. This agrees qualitatively with our
observation that the peak value of the curvature distribution
might be  more sensitive to remnant symmetries than the power law
tail behaviour.

In fact, an ideal experimental realisation of our model may be the
system of two superconduction microwave billiards coupled by an
antenna in a variable way \cite{alt}. Although, in real
experiments of this type the coupling was changed in large
discrete increments, it is in principle possible to change it in a
much more controllable way, and to study level dynamics induced by
such a coupling. We hope that our results may stimulate
experiments of this sort.

\clearpage

%************************************************************

%************************************************************
%\begin{acknowledgments}
We would like to thank  J. Zakrzewski, M. Oxborrow for sending us
copies of their articles \cite{Z-D-curv,quartz-detail} and EPSRC
for the financial support.
%\end{acknowledgments}

%%%%%%%%%%%%%%%%%%%%%%%%%%%%%%%%%%%%%%%%%%%%%%%

\end{document}